# Undeniable Signature Schemes Using Braid Groups

October 14, 2018

Tony Thomas*, Arbind Kumar Lal

Department of Mathematics and Statistics

Indian Institute of Technology Kanpur, Kanpur

Uttar Pradesh, India-208016

{tony,arlal}@iitk.ac.in

**Abstract**

Artin's braid groups have been recently suggested as a new source for public-key cryptography. In this paper we propose the first undeniable signature schemes using the conjugacy problem and the decomposition problem in the braid groups which are believed to be hard problems.

## 1 Introduction

Recently braid groups have been suggested as an alternate platform for doing public-key cryptography. The birthdate of braid group based cryptography can be traced back to the pioneering work of Anshel *et al.* in 1999 [2] and Ko *et al.* in 2000 [16]. Since then, braid groups attracted the attention of many cryptographers due to the fact that, they provide a rich collection of hard problems like the *conjugacy problem, braid decomposition problem* and *root problem* and there are efficient algorithms for parameter generation and group operation [4].

[0]*Research supported by CSIR, New Delhi, under award number: 2-31/97(11)-E.U.II

Since the construction of a Diffie-Hellman type key agreement protocol and a public key encryption scheme by Ko *et al.* in 2000 [16], there have been many attempts to design other cryptographic protocols using braid groups. Positive results in this direction are construction of a pseudorandom number generator by Lee *et al.* in 2001 [19], key agreement protocols by Anshel *et al.* in 2001 [1], an implementation of braid computations by Cha *et al.* in 2001 [4], digital signature schemes by Ko *et al.* in 2002 [15], entity authentication schemes by Sibert *et al.* in 2002 [23] and a provably-secure identification scheme by Kim *et al.* in 2004 [14].

In this paper, we construct some undeniable signature schemes using braid groups. Digital signatures bind signers to the contents of the document they sign. The ability for a third party to verify the validity of a signature is usually seen as the basis for the non-repudiation aspect of digital signatures. The authenticity of a digital signature can be verified by anyone having the public key of the signer. However, this universal verifiability property of digital signatures is not always a desirable property. Such is the case of a signature binding parties to a confidential agreement, or of a signature on documents carrying private or personal information.

Chaum and van Antwerpen [6] introduced the concept of *undeniable signatures* for limiting the ability of third parties to verify the validity of a signature. An undeniable signature, like digital signature depends on the signer's public key as well as on the message signed. Such signatures are characterized by the property that, verification can only be achieved by interacting with the legitimate signer through a *confirmation protocol*. On the other hand, the signer can prove a forgery by engaging in a *denial protocol*. If the signer does not succeed in denying (in particular, if it refuses to cooperate) then the signer remains legally bound to the signature. On the other hand the signer is protected by the fact that his signature cannot be verified by unauthorized third parties without his own cooperation.

Undeniable signatures have got immense real life applications. Almost all the undeniable signature schemes constructed so far have been based on integer factor-ization [10] and discrete logarithm problems [6], [5]. Our work is the first to present undeniable signature schemes based on braid groups or even in any nonabelian group setting. The purpose of this paper is to illustrate the construction of efficient cryptographic protocols based on hard problems in braid groups. Our schemes are based on the conjugacy search problem, multiple simultaneous conjugacy search problem, braid decomposition problem and the multiple simultaneous braid decomposition problem. In Section 2, we briefly review the basics of braid groups. We describe in Section 3 the preliminaries needed for this paper. A simple undeniable signature scheme and a modified one is described in Section 4. A zero-knowledge undeniable signature scheme is given in Section 5. We prove the completeness, soundness and zero-knowledgeness (where ever applicable) of the protocols also. The paper concludes with some general remarks in Section 6.

## 2  An Overview of Braid Groups

In this section, we briefly describe the basics of braid groups, hard problems in braid groups. A good introduction to braid groups is [3] and survey articles on braid cryptography are [17], [7].

### 2.1  Geometric Interpretation of Braids

A braid group $B_n$ is an infinite non-commutative group arising from geometric braids composed of $n$-strands. A braid is obtained by laying down a number of parallel strands and intertwining them so that they run in the same direction. The number of strands is called the braid index. Braids have the following geometric interpretation: an $n$-braid (where $n \in \mathbb{N}$) is a set of disjoint $n$ strands all of which are attached to two horizontal bars at the top and bottom such that each strand always heads downwards as one moves along the strand from top to bottom. Two braids are equivalent if one can be deformed to the other continuously in the set of braids.

Let $B_n$ be the set of all $n$-braids. $B_n$ has a natural group structure. Each $B_n$ is an infinite torsion-free noncommutative group and its elements are called $n$-braids. The multiplication $ab$ of two braids $a$ and $b$ is the braid obtained by positioning $a$ on the top of $b$. The identity $e$ is the braid consisting of $n$ straight vertical strands and the inverse of $a$ is the reflection of $a$ with respect to a horizontal line.

Let $\mathbf{S}_n$ be the symmetric group on $n$ symbols. Given a braid $\alpha$, the strands define a map $p(\alpha)$ from the top set of endpoints to the bottom set of endpoints. In this way we get a homomorphism $p : B_n \to S_n$.

## 2.2 Presentations of Braid Groups

Any braid can be decomposed as a product of simple braids known as *Artin generators* $\sigma_i$, that have a single crossing between the $i^{th}$ strand and the $(i+1)^{th}$ strand with the convention that the $i^{th}$ strand crosses under the $(i+1)^{th}$ strand. The homomorphism, $p$ maps the generator $\sigma_i$ to the transposition $\tau_i$ $(= (i, i+1))$.

For each integer $n \geq 2$, the $n$-braid group $B_n$ has the Artin presentation by generators $\sigma_1, \sigma_2, \ldots, \sigma_{n-1}$ with relations

$$\begin{aligned} \sigma_i \sigma_j &= \sigma_j \sigma_i, \text{ where } |i - j| \geq 2, \text{ and} \\ \sigma_i \sigma_{i+1} \sigma_i &= \sigma_{i+1} \sigma_i \sigma_{i+1}, \text{ for } 1 \leq i \leq n - 2. \end{aligned} \quad (2.2.1)$$

## 2.3 Some Special Classes of Braids

Let $B_n^+$ denote the submonoid of $B_n$ generated by $\{\sigma_1, \ldots, \sigma_{n-1}\}$. Elements of $B_n^+$ are called the *positive braids*. A positive braid is characterized by the fact that at each crossing the string going from left to right undercrosses the string going from right to left.

A positive braid is called *non-repeating* if any two of its strands cross at most once. We denote $D = D_n \subset B_n^+$ to be the set of all non-repeating braids. To each $\pi \in S_n$ we can associate a unique $\alpha \in D_n$ in the following way : for $i = 1, \ldots, n$ connect the upper $i$-th point to the lower $\pi(i)$-th point by a straight line making each *crossing positive*, i.e. the line between $i$ and $\pi(i)$ is under the line between $j$ and $\pi(j)$ if $i < j$. The following lemma says that $p$ restricted to $D_n$ is a bijection.

**Lemma 2.1.** *[9] The homomorphism $p : B_n \to S_n$ restricted to $D_n$ is a bijection.*

Hence non-repeating braids are also known as *permutation braids*. From this lemma it follows that $|D_n| = n!$. In this way we can identify $D_n$ with $S_n$ .

Let $LB_n$ and $RB_n$ be two subgroups of $B_n$ consisting of braids obtained by braiding left $\lfloor \frac{n}{2} \rfloor$ strands and right $n - \lfloor \frac{n}{2} \rfloor$ strands, respectively. That is,

$$LB_n = \langle \sigma_1, \ldots, \sigma_{\lfloor \frac{n}{2} \rfloor - 1} \rangle, \text{ and } RB_n = \langle \sigma_{\lfloor \frac{n}{2} \rfloor + 1}, \ldots, \sigma_{n-1} \rangle.$$

Then we have the commutativity property that for any $\alpha \in LB_n$ and $\beta \in RB_n$, $\alpha\beta = \beta\alpha$. These subgroups of $B_n$ are used in designing various cryptographic protocols.

## 2.4 Canonical Decomposition of Braids

For two words $v$ and $w$ in $B_n$, we say that $v \leq w$, if $w = avb$ for some $a, b \in B_n^+$. Then $\leq$ is a partial order in $B_n$ [9].

The positive braid, $\Delta = (\sigma_1 \ldots \sigma_{n-1})(\sigma_1 \ldots \sigma_{n-2}) \ldots (\sigma_1 \sigma_2) \sigma_1$ is called the *fundamental braid*. A braid satisfying $e \leq A \leq \Delta$ is called a *canonical factor*. There is a bijection between the set of all permutation braids and the set of all canonical factors [9]. Thus a canonical factor can be denoted by the corresponding permutation $\pi \in S_n$. By $\pi_\Delta$, we mean the permutation corresponding to the fundamental braid $\Delta$.

For a positive braid $P$, we say that the decomposition $P = A_0 P_0$ is *left-weighted* if $A_0$ is a canonical factor, $P_0 \geq e$ and $A_0$ has the maximal word length among all such decompositions. A left-weighted decomposition $P = A_0 P_0$ is unique [4]. $A_0$ is called the *maximal head* of $P$. Any braid $x$ can be uniquely decomposed as

$$x = \Delta^u A_1 A_2 \ldots A_k, \text{ where } u \in \mathbb{Z}, A_i \neq e, \Delta, \text{ is a canonical factor} \quad (2.4.1)$$

and the decomposition $A_i A_{i+1}$ is left-weighted for each $1 \leq i \leq k-1$ [4]. This unique decomposition is called the *left canonical form* of $x$ and so it solves the word problem. Since each canonical factor corresponds to a permutation braid, $x$ can be denoted as

$$x = \pi_f^u \pi_1 \pi_2 \ldots \pi_k, \text{ where } \pi_i \neq Identity, \pi_f. \quad (2.4.2)$$

Hence for implementation purposes the braid $x$ can be represented as the tuple $(u, \pi_1, \pi_2, \ldots, \pi_k)$. The integer $u$, denoted by $\inf(x)$ is called the *infimum* of $x$ and the integer $u + k$, denoted by $\sup(x)$ is called the *supremum* of $x$. The *canonical length* of $x$, denoted by $\text{len(x)}$, is given by $k = \sup(x) - \inf(x)$.

## 2.5 Hard Problems in Braid Groups

We use the following hard problems in our signature schemes.

1. **Conjugacy Search Problem (CSP)**

   Let $(x, y) \in B_n \times B_n$, such that $y = a^{-1}xa$, where $a \in B_n$ or some subgroup of $B_n$. The *conjugacy search problem* is to find a $b$ such that $y = b^{-1}xb$.

2. **Multiple Simultaneous Conjugacy Search Problem (MSCSP)**

   Let $(x_1, a^{-1}x_1a), \ldots, (x_r, a^{-1}x_ra) \in B_n \times B_n$ for some $a \in B_n$ or some subgroup of $B_n$. The *multiple simultaneous conjugacy problem* is to find a $b$ such that, $b^{-1}x_1b = a^{-1}x_1a, \ \ldots, \ b^{-1}x_rb = a^{-1}x_ra$.

3. **Braid Decomposition Problem (BDP)**

   Let $(x, y) \in B_n \times B_n$, where $y = a_1xa_2$ for some $(a_1, a_2) \in LB_n \times LB_n$. The *braid decomposition problem* is to find a pair $(b_1, b_2) \in LB_n \times LB_n$ such that $y = b_1xb_2$.

4. **Multiple Simultaneous Braid Decomposition Problem (MSBDP)**

   Let $(x_1, a_1x_1a_2), \ldots, (x_r, a_1x_ra_2) \in B_n \times B_n$ for some $(a_1, a_2) \in LB_n \times LB_n$. The *multiple simultaneous braid decomposition problem* is to find a pair $(b_1, b_2) \in LB_n \times LB_n$ such that, $b_1x_1b_2 = a_1x_1a_2, \ \ldots, \ b_1x_rb_2 = a_1x_ra_2$.

# 3 Preliminaries

In this section, we describe the initial system set up, intractability assumptions, an assumption regarding the cardinalities of certain sets and some notation used in this paper are described.

## 3.1 Initial Setup

The system parameters $n$ and $l$ are chosen to be sufficiently large positive integers and are made public. Let $H : \{0, 1\}^* \to B_n$ and $h : B_n \to \{0, 1\}^k$ be collision free hash functions.

Since the braid group $B_n$ is discrete and infinite, we cannot have a uniform probability distribution on $B_n$. But there are finitely many positive $n$-braids with $l$ canonical factors, we may consider randomness for these braids. Such a braid can be generated by concatenating $l$ random canonical factors.

We fix positive integers $n, l$ as system parameters. Let

$$\begin{aligned} B_n(l) &= \{b \in B_n \mid 0 \leq inf(b) \leq sup(b) \leq l\}, \\ LB_n(l) &= \{b \in LB_n \mid 0 \leq \inf(b) \leq \sup(b) \leq l\} \text{ and} \\ RB_n(l) &= \{b \in RB_n \mid 0 \leq \inf(b) \leq \sup(b) \leq l\}. \end{aligned}$$

Then $|B_n(l)| \leq l(n!)^l$ and so $LB_n(l), RB_n(l)$ and $B_n(l)$ are finite sets. We use the random braid generator given in [4] (which produces random braids in $O(ln)$ time) for generating random braids. Also, we consider uniform probability distribution on these sets.

By, $SRB_p$ we mean some subgroup of $RB_m$, where $p < m - \lfloor \frac{m}{2} \rfloor$ and

$$SRB_p(l) = \{b \in SRB_p \mid 0 \leq \inf(b) \leq \sup(b) \leq l\}.$$

### 3.2 Notations

We use the following notations through out this paper.

- By $a \in_r A$, we mean a random choice of an element $a$ from the set $A$.

- By $P \xrightarrow{Q} V$, we mean $P$ sends $Q$ to $V$.

### 3.3 Intractability Assumptions

We assume that the hard problems *CSP, MSCSP, BDP, MSBDP*, stated in Section 2.5 are intractable in braid groups.

### 3.4 An Assumption on the Cardinality of a Set

We assume that for 'sufficiently large' values of $n$ and $l$ and random choices of $\alpha, \beta, \gamma \in B_n(l)$, $a_1, a_2 \in LB_n(l)$ and $a \in RB_n(l)$, the cardinality of the set $E_a(\beta, \gamma)$

defined by

$$E_a(\beta, \gamma) = \{e \in RB_n(l) : e^{-1}\beta(a_1\alpha a_2)e = a^{-1}\beta(a_1\alpha a_2)a, \ e^{-1}\gamma\alpha e \neq a^{-1}\gamma\alpha a\}$$

is 'sufficiently large'. In this paper, we do not undertake any theoretical or numerical study to check the validity of this assumption. This assumption is rewritten below and will be used in the security analysis of some of our protocols.

**Assumption 3.1.** *Let $n, l$ be 'sufficiently large' positive integers and $\alpha, \beta, \gamma \in_r B_n(l)$, $a_1, a_2 \in_r LB_n(l)$ and $a \in_r RB_n(l)$. Then the cardinality of the set $E_a(\beta, \gamma)$ is bounded below by a nonconstant polynomial function $p(n, l)$ of $n$ and $l$.*

# 4 A Simple Undeniable Signature Scheme

This section describes a simple undeniable signature scheme based on multiple simultaneous braid decomposition problem (MSBDP).

## 4.1 Public and Private Keys

The system is set up by the signer (Alice) in the following manner: Alice chooses random braids $\alpha \in B_n(l)$ and $a_1, a_2 \in LB_n(l)$ and computes $x = a_1\alpha a_2$. She sets her public key as $(\alpha, x)$ and private key as $(a_1, a_2)$.

We shall denote by $PK$ the tuple $(\alpha, x)$ generated as above.

## 4.2 Signature Generation

Suppose that Alice wants to sign a message $m$. She computes $S_m = a_2 y a_1^{-1}$, where $y = H(m)$, giving the output pair $(m, S_m)$. We denote by $SIG(m)$, the set of valid signatures on $m$.

## 4.3 The Confirmation Protocol

Here we present a zero-knowledge confirmation protocol. It is carried out by two players, a prover $(P)$ and a verifier $(V)$. The public input to the protocol are the public key parameters, namely $(\alpha, x) \in PK$ and a pair $(m, \hat{S}_m)$. For the case that $\hat{S}_m$ is a valid signature of $m$, $P$ will be able to convince $V$ of this fact, while if the signature is invalid then no prover (even a computationally unbounded one) will be able to convince $V$ to the contrary except with a negligible probability.

**Signature Confirmation Protocol**

Input : Prover: Secret key $a_1, a_2 \in LB_n(l)$.

Common: Public key $(\alpha, x) \in PK$, $y$ and alleged $\hat{S}_m$.

1. $V$ chooses $a \in_r RB_n(l)$, computes the challenge $Q = a^{-1}(\hat{S}_m x)a$ and $V \xrightarrow{Q} P$.

2. $P$ chooses $b, c \in_r B_n(l)$, computes the response $R = b(a_2^{-1} Q a_2^{-1})c$ and $P \xrightarrow{R} V$.

3. $V \xrightarrow{a} P$.

4. $P$ verifies that $Q = a^{-1}(\hat{S}_m x)a$ and then $P \xrightarrow{(b,c)} V$.

5. $V$ verifies that $R = ba^{-1}(y\alpha)ac$. If it holds then $V$ accepts $\hat{S}_m$ as a valid signature of $P$.

**Theorem 4.1. Confirmation Theorem.** *Let $(\alpha, x) \in PK$.*
**Completeness**: *Given $S_m \in SIG(m)$, if $P$ follows the signature confirmation protocol then $V$ always accepts $S_m$ as a valid signature.*
**Soundness**: *A Cheating prover $P^*$ even computationally unbounded cannot convince $V$ to accept $\hat{S}_m \notin SIG(m)$ with probability greater than $\frac{1}{p(n,l)}$.*
**Zero-knowledgeness**: *The protocol is zero-knowledge, namely on input of a message and its valid signature, any (possibly cheating) verifier $V^*$ interacting with the prover $P$ does not learn any information aside from the validity of the signature.*

*Proof.* **Completeness**: Let $S_m$ be a valid signature. $P$ computes

$$R = b(a_2^{-1} Q a_2^{-1})c = b(a_2^{-1} a^{-1}(\hat{S}_m x) a a_2^{-1})c = ba^{-1}(y\alpha)ac.$$

which $V$ verifies after getting $(b, c)$ from $P$ and accepts the signature as valid. Hence the protocol is complete.

**Soundness**: By Assumption 3.1, there are at least $p(n, l)$ choices for $a \in RB_n(l)$, which give the same value of $Q$ but giving different values of $R$. Hence it is infeasible for a cheating prover $P^*$ to distinguish between these different values of $a$, even if he has infinite computational power. Therefore a cheating prover $P^*$, even computationally unbounded, cannot convince $V$ to accept $\hat{S}_m \notin SIG(m)$ with probability greater than $\frac{1}{p(n,l)}$. Thus the protocol is sound.

**Zero-knowledgeness**: There are two difficulties in the analysis here. Since $B_n$ is an infinite discrete group, we can not have a uniform probability distribution on $B_n$. Also, for the implementation purpose we need to restrict to a finite subset $B_n(l)$ of $B_n$. Unfortunately $B_n$ has no finite nontrivial subgroup. So at some computational stages in the protocol, the elements may fall out of $B_n(l)$. Hence we only give a sketch of the proof here.

For the ease of analysis let us assume that at Step 2, $P$ chooses $(b, c) \in_r B_n^2$ and computes $R \in B_n$. So, we can treat $R$ as a random element of $B_n$. Also since $|B_n(l)| \geq (\lfloor \frac{n-1}{2} \rfloor !)^l$ [16], any random choice of $(b, c) \in B_n(l)^2$ by $P$ in practice makes the response $R$ to appear as a random element. Hence the protocol can be treated as zero-knowledgeable. □

## 4.4 Signature Denial Protocol

The public input to the protocol are the public key parameters, namely $(\alpha, x) \in PK$ and a pair $(m, \hat{S}_m)$. In the case that $\hat{S}_m \notin SIG(m)$, $P$ will be able to convince $V$ of this fact, while if $\hat{S}_m \in SIG(m)$ then no prover (even a computationally unbounded) will be able to convince $V$ that the signature is invalid except with negligible probability.

Let $SRB_{n_1}$ and $SRB_{n_2}$ be two subgroups of $RB_n$ consisting of braids obtained by braiding left $n_1$ strands and right $n_2$ strands, respectively (where $n_1 + n_2 = n - \lfloor \frac{n}{2} \rfloor$). Then we have the commutativity property that for any $\sigma \in SRB_{n_1}$ and $\tau \in SRB_{n_2}$, $\sigma\tau = \tau\sigma$.

**The Denial Protocol**

Input : Prover: Secret key $a_1, a_2 \in LB_n(l)$.

Common: Public key $(\alpha, x) \in PK$, $y$ and alleged $\hat{S}_m$.

1. $V$ chooses $a \in_r SRB_{n_1}(l)$, $b \in_r SRB_{n_2}(l)$, computes $Q_1 = a^{-1}\hat{S}_m xa$, $Q_2 = b^{-1}\hat{S}_m xb$ and $V \stackrel{(Q_1,Q_2)}{\longrightarrow} P$.

2. $P$ computes the responses $R_1 = a_2^{-1}Q_1 a_2^{-1}$, $R_2 = a_2^{-1}Q_2 a_2^{-1}$ and $P \stackrel{(R_1,R_2)}{\longrightarrow} V$.

3. $V$ verifies $b^{-1}R_1 b = a^{-1}R_2 a$. If equality holds $V$ accepts $\hat{S}_m$ as an invalid signature, else $P$ is answering improperly.

**Theorem 4.2. Denial Theorem** *Let $(\alpha, x) \in PK$*

**Completeness**: *Suppose that $\hat{S}_m \notin SIG(m)$. If $P$ and $V$ follow the protocol, then $V$ always accepts that $\hat{S}_m$ is not a valid signature of $m$.*

**Soundness**: *Suppose that $\hat{S}_m \in SIG(m)$. Then a cheating prover, even computationally unbounded, cannot convince $V$ to reject the signature with probability greater than $\frac{1}{p(n,l)}$.*

*Proof.* **Completeness**: Assume that $\hat{S}_m \notin SIG(m)$. We have,

$$R_1 = a_2^{-1}a^{-1}\hat{S}_m xaa_2^{-1} = a^{-1}a_2^{-1}\hat{S}_m a_1 \alpha a \quad \text{and}$$
$$R_2 = a_2^{-1}b^{-1}\hat{S}_m xba_2^{-1} = b^{-1}a_2^{-1}\hat{S}_m a_1 \alpha b.$$

As $ab = ba$, we get,

$$b^{-1}R_1 b = a^{-1}R_2 a = a^{-1}b^{-1}(a_2 \hat{S}_m a_1 \alpha)ba.$$

Hence the protocol is complete.

**Soundness**: Assume that $\hat{S}_m \in SIG(m)$. Let $R_1$ and $R_2$ be the responses given by $P^*$ in the protocol. Let if possible, $b^{-1}R_1 b = a^{-1}R_2 a$. Then

$$R_2 = a(b^{-1}R_1 b)a^{-1} = a\beta a^{-1}, \text{ where } \beta = b^{-1}R_1 b.$$

In the worst case, we may regard $\beta$ as a known constant for $P$ when he tries to determine $R_2$. But then the ability to determine $R_2$ amounts to the establishment of an invalid signature, which contradicts Theorem 4.1 (soundness of the confirmation protocol). Hence the protocol is sound. □

## 4.5 A Blackmailing Attack

In this subsection, we show that any non zero-knowledge version of the confirmation protocol is susceptible to a blackmailing attack. That is, if the prover does not check the challenge $Q$ of the verifier, he is susceptible to a black mailing attack. This type of attack for the discrete logarithm based undeniable signature schemes was suggested by M. Jakobsson [12]. M. Jakobsson noted that the protocol in [6] has the following weakness that Alice proving the correctness of her signatures never knows what signature is being verified. Using these weaknesses of the undeniable signatures he showed that how an adversary can blackmail a signer.

For illustrating this attack in our case, we consider the following non-zero knowledge version of the confirmation protocol given in Section 4.3.

**Signature Confirmation Protocol**

Input : Prover: Secret key $a_1, a_2 \in LB_n(l)$.

Common: Public key $(\alpha, x) \in PK$, $y$ and alleged $\hat{S}_m$.

1. $V$ chooses $a \in_r RB_n(l)$, computes the challenge $Q = a^{-1}(\hat{S}_m x)a$ and $V \xrightarrow{Q} P$.

2. $P$ computes the response $R = a_2^{-1} Q a_2^{-1}$ and $P \xrightarrow{R} V$.

3. $V$ verifies that $R = a^{-1}(y\alpha)a$. If it holds, $V$ accepts $\hat{S}_m$ as a valid signature of $P$.

Now, suppose that Eve has found out that $(m, S_m)$ belongs to Alice (Eve might be Bob itself). Now we will show, how Eve can convince $k$ entities $E_1, E_2, \ldots, E_k$ that the signature pair belongs to Alice.

Let $SRB_{n_0}, SRB_{n_1}, \ldots, SRB_{n_k}$ be $k+1$ subgroups of the $n$-braid group $RB_n$, where $n - \lfloor \frac{n}{2} \rfloor = n_0 + n_1 + \ldots + n_k$, for some appropriate positive integers $n_0, n_1, \ldots, n_k$. Each $SRB_{n_i}$ is the subgroup of $RB_n$ consisting of braids made by braiding $n_i$-strands from the left among $n - \lfloor \frac{n}{2} \rfloor$ strands with the order $n_0, n_1, \ldots, n_k$. Let $n_{-1} = 1$. Then for $i = 0, 1, \ldots, k$,

$$SRB_{n_i} = \langle \sigma_{l_i+1}, \sigma_{l_i+2}, \ldots, \sigma_{l_i+n_i-1} \rangle, \quad \text{where} \quad l_i = \lfloor \frac{n}{2} \rfloor + \sum_{j=0}^{i-1} n_j$$

and we have the mutual commutativity property that for any $\alpha_i \in SRB_{n_i}$ and $\alpha_j \in SRB_{n_j}$ with $i \neq j$, $\alpha_i \alpha_j = \alpha_j \alpha_i$.

**The Protocol**

1. *Eve* asks each $\{E_i\}_{i=1}^k$ to choose secret braids $a_i \in SRB_{n_i}(l)$.

2. *Eve* chooses a secret braid $a_0 \in RB_{n_0}(l)$ and computes $Q_0 = a_0^{-1} S_m x a_0$.

3. $Eve \xrightarrow{Q_0} E_1 \xrightarrow{Q_1} E_2 \xrightarrow{Q_2} \ldots \xrightarrow{Q_{k-1}} E_k \xrightarrow{Q_k} Eve$, where $Q_i = a_i^{-1} Q_{i-1} a_i$, for $i = 1, \ldots k$.

4. *Eve* convinces *Alice* to engage in a confirmation protocol for a message pair $(\hat{m}, S_{\hat{m}})$.

5. $Eve \xrightarrow{Q_k} Alice$.

6. *Alice* computes the response $R = a_2^{-1} Q_k a_2^{-1}$ and $Alice \xrightarrow{R} Eve$.

7. $Eve \xrightarrow{(Q_k, R, m)} E_i$, for $i = 1, \ldots k$.

8. $E_i \xrightarrow{a_i} Eve$, for $i = 1, \ldots k$.

9. *Eve* computes $a = \prod_{i=0}^{k} a_i$ and $Eve \xrightarrow{a} E_i$, for $i = 1, \ldots k$.

10. Each $E_i$ checks whether $Q_k = a^{-1} S_m x a$ and $R = a^{-1} y \alpha a$. If it holds they will be convinced that *Alice* signed the message $m$.

**Theorem 4.3.** *If Eve sends out $(a, R, Q_k, m)$ to $\{E_i\}_{i=1}^k$, each one of them will be able to convince himself that the signature belongs to Alice.*

*Proof.* By Assumption 3.1, it follows that *Eve* can not get $\{a_i\}_{i=1}^k$ from $Q_k$ before committing $(Q_k, R, m)$ to $\{E_i\}_{i=1}^k$, since each $a_i$ is a random braid chosen and kept secret by $E_i$. By checking $Q_k = a^{-1} S_m x a$, each $E_i$ will be convinced that Eve has not cheated them by forming the challenge $Q_0$ corresponding to some other signer. □

## 4.6 Blinding

The above signature scheme is a deterministic signature scheme whose security is based on the hardness of MSBDP. MSBDP may become easier as the number of available braid pairs increases. In our case, the parameter $r$ in MSBDP is the number of messages signed. So to make the scheme more secure, we may modify the scheme by blinding the signatures using random braids. The modified scheme can be described as follows.

**Signature Generation**

To generate a signature on a message $m$ the signer proceeds in the following way.

1. Signer chooses $r \in_r LB_n(l)$.

2. Computes $S_m = ra_2ya_1^{-1}$, where $y = H(m)$.

3. Outputs the signature $(m, S_m)$.

The confirmation and denial protocols are exactly similar to the protocols given in the earlier case. A non zero-knowledge version of the confirmation protocol is given below for an illustration.

**The Confirmation Protocol**

Input : Prover: Secret key $a_1, a_2 \in LB_n(l)$, and blinding factor $r \in_r LB_n(l)$.
 Common: Public key $(\alpha, x) \in PK$, $y$, and alleged $\hat{S}_m$.

1. $V$ chooses $a \in RB_n(l)$ computes the challenge $Q = a^{-1}\hat{S}_m xa$ and $V \xrightarrow{Q} P$.

2. $P$ computes the response $R = a_2^{-1}r^{-1}Qa_2^{-1}$ and $P \xrightarrow{R} V$.

3. $V$ verifies that $R = a^{-1}y\alpha a$. If equality holds then $V$ accepts $\hat{S}_m$ as the signature of $P$.

**Remark 4.1.** *We can easily see that the confirmation theorem and the denial theorem hold in this case also.*

# 5 A Zero-knowledge Undeniable Signature Scheme

In this section, we describe an undeniable signature scheme in which the denial protocol is also zero-knowledge.

## 5.1 Public and Private Keys

The system is set up by the signer (Alice) in the following manner: Alice chooses $\alpha \in_r B_n(l)$ and $a_1, a_2 \in_r LB_n(l)$ and computes $x = a_1 \alpha a_2$. She sets her public key as $(\alpha, x)$ and the private key as $(a_1, a_2)$.

We shall denote by $PK$, the tuples $(\alpha, x)$ generated as above.

## 5.2 Signature Generation

Suppose that Alice wants to sign a message $m$. She computes $S_m = a_1 y a_1^{-1}$, where $y = H(m)$, giving the output pair $(m, S_m)$.

We denote by $SIG(m)$, the set of valid signatures on $m$.

## 5.3 The Confirmation Protocol

The confirmation protocol in this case is exactly similar to the protocol given in Section 4.3.

## 5.4 Denial Protocol

Here we describe a zero-knowledge denial protocol. The public input to the protocol are the public key parameters, namely $(\alpha, x) \in PK$ and a pair $(m, \hat{S}_m)$.

In this protocol, we use a zero-knowledge commitment function called *blob*. $blob(r, t)$ perfectly hides the value of $t$ as long as $r$ is secret and once the value of $r$ is revealed one can open the *blob* and get the value of $t$.

**Signature Denial Protocol**

Input : Prover : Secret key $a_1, a_2 \in LB_n(l)$.

        Common: Public key $(\alpha, x) \in PK$, $y$ and alleged $\hat{S}_m$.

1. $V$ chooses $a \in_r RB_n(l)$ and $t \in_r \{1, 2, \ldots, k\}$, computes
$$Q = (y^t a^{-1} \alpha a, \hat{S}_m^t a^{-1} x a) = (Q_1, Q_2) \text{ and } V \xrightarrow{Q} P.$$

2. $P$ computes $t$ by trial and error using $Q_2/a_1 Q_1 a_2 = (\hat{S}_m/s^{-1} y s)^t$.

   Also, $P$ chooses $r$ randomly and $P \xrightarrow{blob(r,t)} V$.

3. $V \xrightarrow{a} P$.

4. $P$ checks the value of $Q$ using $a$ and then $P \xrightarrow{r} V$.

5. $V$ opens the *blob* using the value of $r$ and checks the value of $t$. If the value of $t$ committed by $P$ is correct, then $V$ accepts that $\hat{S}_m$ is not a valid signature of $P$.

**Remark 5.1.** *The value of $k$ in Step 1 above depends on the computing power of the prover and the verifier. If the prover has low computing power, the value of $k$ can be chosen to be small but then the protocol needs to be repeated.*

**Theorem 5.1. Denial Theorem** *Let $(\alpha, x) \in PK$.*

**Completeness**: *If $\hat{S}_m \notin SIG(m)$ and if $P$ and $V$ follow the protocol, then $V$ always accepts that $\hat{S}_m$ is not a valid signature of $m$.*

**Soundness**: *Assuming that $\hat{S}_m \in SIG(m)$, then a cheating prover, even computationally unbounded, can not convince $V$ to reject $\hat{S}_m$ with probability greater than $1/k$.*

**Zero-knowledgeness**: *The protocol is zero-knowledge, namely, on input of a message and a non valid signature, any (possibly cheating) verifier $V^*$ interacting with the prover $P$ does not learn any information aside from the fact that $\hat{S}_m$ is in fact not a valid signature for the message $m$.*

*Proof.* **Completeness**: Upon receiving $Q$ from $V$, $P$ computes

$$\begin{aligned}
Q_2/a_1 Q_1 a_2 &= ((\hat{S}_m)^t a^{-1} x a)/(a_1 y^t a_1^{-1})(a_1 a^{-1} \alpha a a_2) \\
&= ((\hat{S}_m)^t a^{-1} x a)/(a_1 y^t a_1^{-1})(a^{-1} x a) \\
&= (\hat{S}_m/a_1 y a_1^{-1})^t \neq e.
\end{aligned}$$

Since $(Q_2/a_1 Q_1 a_2)$ and $(\hat{S}_m/a_1 y a_1^{-1})$ are known to $P$, $P$ can compute the value of $t$ by trial and error as the value of $t$ is small.

**Soundness**: $a$ hides $t$ in the challenge $Q$. Since the value committed to by the blob cannot be changed, $P$'s best strategy is to guess the value of $t$, and there are $k$ choices for $t$. Hence the protocol is sound.

**Zero-knowledgeness**: This follows immediately from the zero-knowledge commitment of the *blob*. □

**Remark 5.2.** *The above signature scheme is a deterministic signature scheme whose security is based on the hardness of MSCSP. As in the case of MSBDP, MSCSP may become easier as the number of available conjugate pairs increases. Hence the scheme may be made more secure by blinding the signatures using random braids as described in Section 4.*

# 6 Concluding Remarks

In this paper, we constructed some undeniable signature schemes using braid groups. Some of these schemes enjoy the zero-knowledge property. We used braid groups for the first time for designing undeniable signatures. The security of our schemes are based on the hardness of CSP, MSCSP, BDP and MSBDP. One can also explore the possibility of employing other hard problems for designing these protocols.

In Step 1 of the confirmation protocol given in Section 4.3, $V$ may compute the challenge $Q$ as $a\hat{S}_m xb$ or $a^{-1}\hat{S}_m ab^{-1}xb$ or $a\hat{S}_m bxc$ instead of $a^{-1}\hat{S}_m xa$, where $a, b, c \in RB_n(l)$. The advantage with this modification is that there can be more choices, for $a, b, c \in RB_n(l)$ which give the same value of $Q$ but giving different values of $R$. This makes the task of a cheating prover $P^*$ to guess the value of $R$ harder, which in turn makes the probability for a cheating prover $P^*$ to convince $V$ to accept $\hat{S}_m \notin SIG(m)$ smaller. The denial protocol given in Section 4.4 can also be similarly modified.

In this paper we have not carried out any investigation regarding the validity of the Assumption 3.1. We leave this problem for future investigation. Getting a theoretical justification for this assumption appears to be too hard. However, numerical experiments might throw some light on this assumption.

There are many desirable features for a good undeniable signatures like *con-*

*vertibility* (the possibility to transform undeniable signatures into regular ones), *delegation* (enabling selected third parties to confirm/deny signatures but not to sign). We have not considered these problems in this paper. Hence we hope that this study will motivate further research on digital signature schemes based on braid group as well as other nonabelian groups.

The birth of braid cryptography has stimulated the search for other exotic mathematical structures for doing public-key cryptography. The public-key cryptography has been treated under the head of number theory and finite fields only. With the birth of braid cryptography a broader perspective on public-key cryptography has emerged. People have started looking at other nonabelian groups [25], [24], [20], [11] and combinatorial groups [22], [21] for building public-key cryptosystems. Hence we hope that this study will further stimulate the search for other mathematical structures as a better alternative to the number theoretic and discrete log based cryptosystems.